# Electromagnetic Induction and the Conservation of Momentum in the Spiral Paradox.

## Albert Serra -Valls

**Departamento de Física, Facultad de Ciencias, Universidad de Los Andes, Venezuela.**

To the memory of my professor and late friend Salvador Velayos,

who once said "The spiral disturbs me"

The inversion of cause and effect in the classic description of electromagnetism, gives rise to a conceptual error which is at the bottom of many paradoxes and exceptions. At present, the curious fact that unipolar induction or the Faraday Disc constitutes an exception to the Faraday induction law is generally accepted. When we establish the correct cause and effect relationship a close connection appears between mechanics and electromagnetism, as does a new induction law for which paradoxes or exceptions do not occur. Difficulties in interpreting the Faraday Disc derive directly from Faraday's Induction Law and the equation that defines and measures magnetic induction. The electromagnetic force and the unipolar torque generated in the Faraday Disc depend on the shape of the circuit which connects to the disc, giving rise to an "absolute-relative" duality of the emf and unipolar torque. This gives rise to different interpretations. Analogy with mechanics suggests this duality derives from the twofold inert and gravitational nature of the electromagnetic mass. Some paradoxical experiments in unipolar induction involving the unique geometry of the spiral are described demonstrating this duality and the inversion of cause and effect. The emf and torque of the Faraday Disc and the conducting spiral is due to the continuous variation of the electromagnetic angular moment of the continuous current. This experiments confirm the Lorentz Force and invalidate Faraday's Induction Law. They show how in a closed circuit emf and unipolar torque are not produced by the variation in magnetic flux, which is constant, but by two variations in the electromagnetic angular moment. The three possible ways of varying the electromagnetic angular moment generated by the circulation of the charges gives rise to the different forms of electromagnetic induction.

# Introduction.

## 1. Unipolar or acyclic induction.

Possibly, there has been no simpler, more curious and polemical experiment since the beginnings of electromagnetism than Faraday's rotating magnet and disc. For their simplicity and beauty they have always attracted the attention of the physicist.

According to Poincaré "The most curious electrodynamics experiments are those where a continuous rotation takes place, called unipolar induction experiments."[1]

Einstein, in his first paper "On the electrodynamics of moving bodies," states that: "It is known that Maxwell's electrodynamics –as usually understood at the present time– when applied to moving bodies, leads to asymmetries which do not appear to be inherent in the phenomena". "Furthermore it is clear that the asymmetry mentioned in the introduction as arising when we consider the currents produced by the relative motion of a magnet and a conductor, now disappears. *Moreover, questions as to the "seat" of electrodynamic electromotive forces (unipolar machines) now have no point.*"[2]

It would seem the Faraday disc contributed to the development of the Theory of Relativity.

When studying unipolar induction back 1961, and finding the conducting spiral to be a universal unipolar generator I imagined that this must have been known since the beginnings of electromagnetism. In that year I had begun my Ph. D. course in Physics at Grenoble University and found to my surprise that the conducting spiral was unknown to my professors of electromagnetism. They suggested that I choose this for a second subject for my doctoral thesis[3]. It turned out to be very polemical, for as is well-known, unipolar induction continues to be the object of discussions and publications. On completing my thesis, the Board of Examiners recommended my second subject for publication; something I was only able to do years later, for in the opinion of the journal's referee the conducting spiral was but a "mind experiment" and couldn't possibly revolve. Only on checking the experiment (presumably), was the article accepted. This publication[4] had involved considerable difficulties and scarce attention. To start with, I - the supposed discoverer - had failed to grasp the significance of the spiral. Curiously, this experiment, as straightforward and beautiful as the Faraday Disc, is just as paradoxical. Twenty-seven years after publishing my article I began my studies of unipolar induction anew with a series of experiments on conducting spirals which led me to a new understanding of electromagnetic induction, the Faraday Disc and the conducting spiral itself, establishing a new analogy between mechanics and electromagnetism. In November 1998, I attempted to publish these findings in the same journal which in 1970 had published my first article, only to have it rejected out of hand by the editor who alleged "articles announcing new theoretical results or experiments are not accepted in this journal". Maybe he should have added: especially if they come from an unknown third-world Physicist, for this publication continues to carry articles on Faraday's Induction Law and the Lorentz Force [5, 6, 7, 8], all of which deal with the old question as to how and where emf is generated in the Faraday Disc. Regarding the substance of the matter, some authors are of the opinion that the revolving magnet and the Faraday Disc are exceptions to Faraday's Induction Law or flux rule[9], and assure us that unipolar induction is due to the Lorentz Force, others deny any exceptions[10], and still others see exceptions to the Lorentz Force[11].

The difficulties in understanding the Faraday Disc derive from Faraday's Induction Law and the equation F = il x B, which defines B and allows it to be measured. This assumes that magnetic induction B, generated by the circuit to which the segment l belongs, is negligible with regard to B. The emf and torque generated in the Faraday Disc depend on the shape of the circuit that connects the disc, giving rise to an "absolute – relative" duality of emf and Lorentz Force, which in turn, occasions different interpretations. This duality becomes much more evident in the conducting spiral and when the symmetry of the Faraday Disc is enhanced.

Some paradoxical experiments in unipolar induction which make use of the unique geometry of the spiral are described in this article. These experiments show that the paradoxes and discrepancies that arise with unipolar induction are resolved when the following analogies between mechanics and electromagnetism are established:

a) Charges, in the same way as mass, have a dual nature, inert and gravitational, in each of these pairs neither element is independent of the other.

b) In electromagnetic interaction among charges, both mechanical and electromagnetic angular moments are conserved.

c) Electromagnetic induction is due to the variation and conservation of the angular moments of mass and charge.

d) The possible ways of varying the electromagnetic angular moment of a current in a circuit correspond to the forms of electromagnetic induction.

e) The deformation of a circuit by electromagnetic forces tends to diminish the rate of change of the

electromagnetic angular moment of the current's charges, i.e. it will tend to conserve the angular moment.

The circulation of the charges of the continuous current in a Faraday Disc, as also in a conducting spiral, generates a continuous rate of change of angular electromagnetic moment and angular moment of matter, this works in the same way as an electrodynamic turbine. Due to the coexistence and conservation of the angular moment of the electromagnetic field and of matter, in all closed circuits there are always two equal and opposite variations of the angular moment generated.

In closed circuits, constant emf is not produced by the variation in magnetic flux, which is constant, but by two variations in the electromagnetic angular moment.

This means the new induction law will be $\varepsilon = -dL/dt$ $d\varepsilon/dt = -d\phi/dt$ in which L is the electromagnetic angular momentum and $\phi$ is the magnetic flux density.

According to this new induction law, unipolar induction is a consequence and not an exception.

The generation and variation of the angular moments of the electromagnetic field and of matter, occur through the normal constraint forces acting along the path of the charges in the conductors. These constraint forces are not explicit in Maxwell's equations. However, without these forces it would not be possible to generate or measure electric or magnetic field.

The conducting spiral allows us to see that unipolar induction is produced by a vortex of charges, confirming the Lorentz Force and invalidating Faraday's Induction Law, furthermore it allows us to see the true origin of electromagnetic induction and its dual nature. In the conducting spiral, an inversion of cause and effect in the description of electromagnetism also becomes evident.

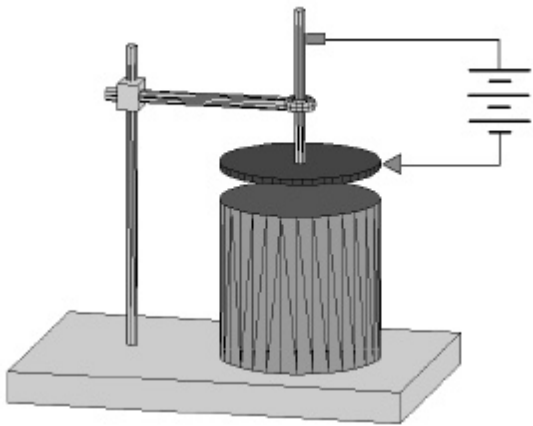
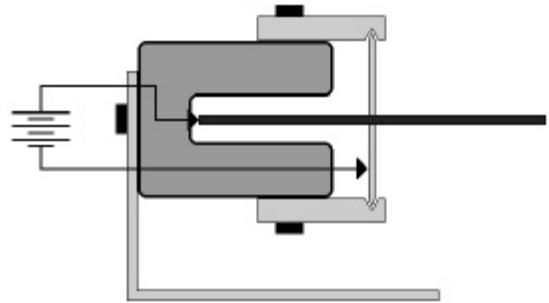

| Fig. 1.a. The Faraday disc | Fig. 1.b. The Barlow wheel |
|---|---|
| • The magnetic field on the disc has rotational symmetry. | • The magnetic field on the disc doesn't have rotational symmetry. |
| • The rotation axis of the disc and the rotational symmetry axis of the magnetic field of the magnet coincide. | • The magnetic field is parallel to the rotation axis but is not symmetric with respect to this axis. |
| • Eddy currents are not induced on the disc. | • Eddy currents are induced on the disc. |
| • In open circuit the magnet does not brake the disc. | • In open circuit the magnet brakes the disc. |
| • The magnet is not the stator and can rotate together with the disc. | • The magnet is the stator and cannot rotate with the disc. |
| • The disc and the magnet don't constitute the whole generator. | • The disc and the magnet constitute the whole generator. |

## 2. Exceptions to Faraday's Law or the "Flux Rule".

Possibly owing to the polemical nature of the subject, even the best texbooks[12] provide only a very superficial discussion of the Faraday Disc, and do not mention experimental facts which are fundamental to the comprehension of unipolar induction. An example of this are the differences that exist between the Faraday Disc and the Barlow Wheel as shown in Fig. 1.

The rotation of a cylindrical magnet around its axis does not induce, nor is it braked by the presence of a conductor. . This is due to the coincidence of the rotation axis and the rotational symmetry axis of the magnetic field of the cylindrical magnet. This fact differentiates Faraday's Disc from other types of generators

In the Faraday Disc, the circuit that closes the current of the disc, mysteriously, is not usually taken into account in the description of the experiment; "the dark side of the force" is the stator.

Here it is easy to see that the field lines are closed ($\nabla \cdot B = 0$), increasing the diameter of the disc; the induction lines will cross the disc twice, annulling the torque and the emf, Fig. 2.

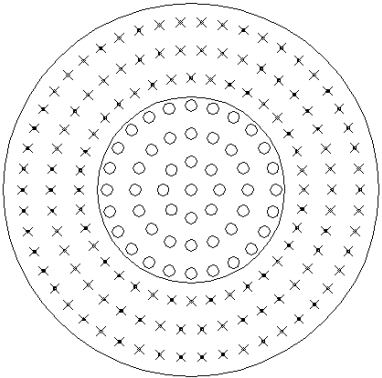

*Fig. 2. The distribution of the magnetic induction lines in the Faraday Disc. emf and torque are maximum when brush takes place at the circumference that separates the two directions of the induction lines.*

The brush on the disc should be placed on the circumference where the induction lines are inverted. The induction lines B should cross the disc once only.

This fact is not taken into account in the description of the Faraday disc made by R. Feynman in his objection to the flux rule in the Faraday Disc[13] The experiment described is a mixture of the Faraday disc and the Barlow wheel, and it is doubtful whether it would work, Fig. 3

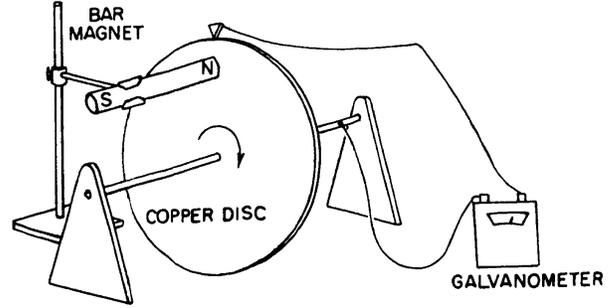

*Fig. 3. Description of the Faraday Disc according to The Feynman Lectures on Physics Vol. II, P. 17-2. "When the disc rotates there is an emf from v x B, but with no change in the linked flux."*

In the first place, most of the lines of the magnet cut across the disc twice, and this hinders their functioning as generator. In the second place, the magnet's axis doesn't coincide with the disc's rotation axis in which the Eddy current or Foucault current are induced, as is the case with the Barlow Wheel, due to the flux change. In this experiment the idea is to describe the Faraday Disc "in which no rate of change of flux occurs". When the rotation axis of the disc and the rotational symmetry axis of the magnet's field don't coincide, Eddy currents are induced in the disc. This fact is mentioned by Scorgie at the end of his article.[14]

Some physicists, in order to safeguard the Flux Rule or the Faraday Law in unipolar induction, take the rate of flux swept by a radius of the Faraday Disc, as a true variation of the linked flux. "The emf may also be calculated by using the Faraday law. The only problem is how to choose the circuit through which the changing flux is to be calculated"[15].

The only problem actually, resides in that *constant* changing flux is not detected by a curl-meter.

As is well known, the calculation of the emf induced on the Faraday Disc, applying the Lorentz force or the swept flux rule gives the same result.

$$\text{emf} = 1/2 \, B \, \omega \, R^2 \qquad (1)$$

According to Faraday's induction law $\varepsilon = -d\phi/dt$, the generation of this constant emf implies a constant rate of change of magnetic flux and its indefinite growth in the circuit. For this reason, some authors believe that the Faraday Disc constitutes an exception to Faraday's Law or the flux rule. Further on, we shall see that according to Feynman, Galili and Kaplan[16], **the density magnetic flux** does not vary and this is a necessary condition for the generation of a constant emf.

## 3. The Faraday Disc and the absolute - relative duality.

In the Faraday Disc the relative rotation between disc and magnet is irrelevant. The variation in magnetic induction in the magnet due to its rotation (Barnett Effect) is negligible, for this reason unipolar emf does not depend on relative angular velocity between disc and magnet. Rotation occurs relative to the magnetic field generated by the magnet, but the rotation of the magnet on its axis does not change the magnetic field.[17] That is to say, that the magnetic field of the cylindrical magnet has rotational symmetry and does not vary when the magnet revolves round the symmetry axis. Thus, there is no conceptual difference at all between the disc and Faraday's rotating magnet as some authors believe.[18, 19]

To facilitate the analysis of unipolar induction, we can give the Faraday Disc a larger rotational symmetry. The Faraday Disc and the rest of the circuit can be replaced by two hemispheres whose rotation axis coincides with the axis of the cylindrical magnet enclosed inside it. Fig. 4.

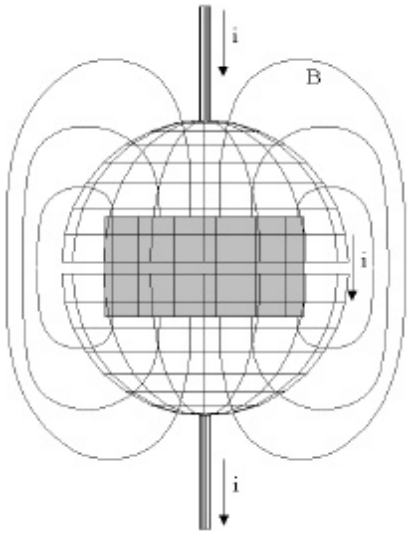

*Fig. 4. Cylindrical magnet contained inside a conducting* sphere, *whose hemispheres can rotate independently around their axis coinciding with the axis of rotational symmetry of the magnet's field.*

If we split the cylindrical magnet in two equal parts at its equator and fix one of the halves to each hemisphere, nothing will change in the experiment, but it will be made easier to carry out.

The identical hemispheres constitute the disc or rotor and the stator respectively, in the Faraday experiment. If the conducting surface of the sphere is traveled over by a continuous current that enters, for example, from the North Pole and goes out from the South, an equal and opposite torque is generated on the hemispheres that makes them rotate in a opposite directions, when allowed to slide on the equator. The rotor and the stator are indistinguishable. Reciprocally, if one hemisphere is made to rotate on top of the other, an emf is generated between the poles.

We may ask whether this emf is produced by the absolute rotation of one of the hemispheres in the magnetic field, as prescribed by the Lorentz Force, and completely independently of the existence of the other hemisphere; or whether the emf is produced by the relative rotation of the two hemispheres in the magnetic field. We may observe that this constitutes an absolute-relative duality which is totally equivalent and indistinguishable in a closed circuit. This type of Faraday Disc, whose rotor and stator are identical, makes this duality easier to apprehend.

Rotation in the magnetic field of the two hemispheres conjointly as if it were a single sphere, does not generate any emf between the poles. This may be attributed to the absence of relative movement, or that emfs are equal and cancel each other out. When both hemispheres rotate together as a single sphere in a magnetic field, according to the theory of relativity or to the Lorentz Force, an electric field is generated which produces an emf $\int E \cdot dl$ between the pole and the equator and which is equal in the two hemispheres and which cancel each other out. To measure this unipolar emf between the pole and the equetor we required a stationary conductor (hemisphere) and a sliding contact. This is equivalent to producing a rotation between the hemispheres.

When the two hemispheres rotate with the same angular velocity but in opposite directions, according to the theory of relativity or the Lorentz Force, an equal and opposite emf is generated which doubles the emf between the poles while the relative angular velocity between the hemispheres is also doubled. It can be seen that the two interpretations are evidently indistinguishable and constitute an "absolute–relative" duality.

Are the emfs in the twin hemispheres independent of each other?

We shall see further on, that the need for the conservation of the electromagnetic angular moment of the current in a closed circuit causes the emfs generated in both hemispheres or circuit parts in relative movement not to be independent.

The statement "we need only consider a radial "rod" in the rotating disc"[20] supposes that the emf $1/2B\omega R^2$ all along the rod will not alter on closing the circuit. In order to calculate unipolar emf, the rest of the circuit is quite unnecessary, however it is absolutely essential for generating and measuring the emf and constant current.

In my opinion this is the crux of the matter. When calculating the emf produced by induction in a section of open circuit by using the Lorentz Force, as one would calculate the difference in potential in an electrostatic field ∫E•dl, we forget that the electric field lines, generated by induction, are closed, analogous to those of a magnetic field. The emf is calculated in a section of an open circuit, but is generated and measured in a closed circuit. As will be seen further on, the emf depends on the shape of the conductor that closes the circuit, this occurs when it forms part of the generator, as is the case with the Faraday Disc.

Emf is attributed to the rotating hemisphere and is calculated on it. Supposedly, the emf induced in the stationary hemisphere is zero.

According to the Lorentz Force, the emf is located in the rotating hemisphere. This contradicts the statement underlined in the introduction **"Moreover, questions as to the "seat" of electrodynamic electromotive forces (unipolar machines) now have no point"[21].**

**It is an experimental fact that when current circulates between the poles or is generated by rotating a hemisphere, two equal and opposite torques are generated. These torques are attributed to the Lorentz force.**

This suggests, according to the symmetry involved, that in a **closed circuit**, emf should also be generated in both hemispheres. Which is to say, that in a closed circuit there should not be a single unipolar emf located in a hemisphere that rotates in relation to the magnetic field, but that emf is located in both hemispheres, as occurs with the torques. In a closed circuit the emf generated is the result of two equal and opposite emfs which do not cancel each other out. These are located in each one of the two parts of the circuit which are in movement, each one in relation to the other. The emf generated between the centers of the discs or the poles of the hemispheres depends on the difference in angular rotation velocity of each hemisphere or what comes to the same, the relative velocity between them. This gives rise to an absolute-relative duality in unipolar induction.

**Even though we calculate and explain the generation of an emf in a conductor (disc) that rotates in a magnetic field, it is not possible to generate a constant emf and current unless we complete the circuit and produce a relative movement between the parts.**

In the following sections, we go on to describe some experiments in unipolar induction involving spiral-shaped conductors without any magnetic field other than that generated by the spiral itself. These experiments confirm and make clear beyond any doubt, this absolute-relative duality, showing that due to the conservation of the electromagnetic angular moment of the current, unipolar emf and constant current are the result of two equal and opposite emfs.

Furthermore, these experiments provide an insight into the inversion of cause and effect in electromagnetic induction and the Lorentz Force, because they make it possible to see the part played by the variation and conservation of the electromagnetic angular moment generated by the circulation of the charges in the circuit. All these experimental findings must be put down to the coexistence and inviolability of the conservation of the mechanical and electromagnetic angular moments.

## 4. The Symmetrical or Twin Faraday Disc.

This experiment with the double, symmetrical Faraday Disc, was carried out by making use of two identical rare earth magnets each fixed to one of two identical iron discs. These discs have a circular channel made near their borders in which some brass balls are inserted allowing the iron discs to rotate round their axis. The twin discs are connected electrically through these brass balls. A conducting disc (Faraday Disc) is placed between the magnet and the iron disc. The discs are connected electrically through their rims. In this way a radial current enters via the central axis of one of the conducting discs and leaves via the center of the other one. See Fig. 5. This experiment requires two concentric Faraday Discs connected electrically throughout their rims. This twin Faraday Disc has a total rotational symmetry (rotor-stator) of its magnetic induction and radial currents. The point of contact on the rim is replaced by a circumferential contact.

To emphasize the Angular Moment Conservation Principle, and to ascertain that we have here a complete generator (the feed wires do not form part of this motor, as was the case with the Faraday Disc), rotor and stator are identical and the components are mounted so that both parts can rotate in relation to the laboratory. Fig. 5.

Electromagnetic Induction and the Conservation of Momentum in the Spiral Paradox

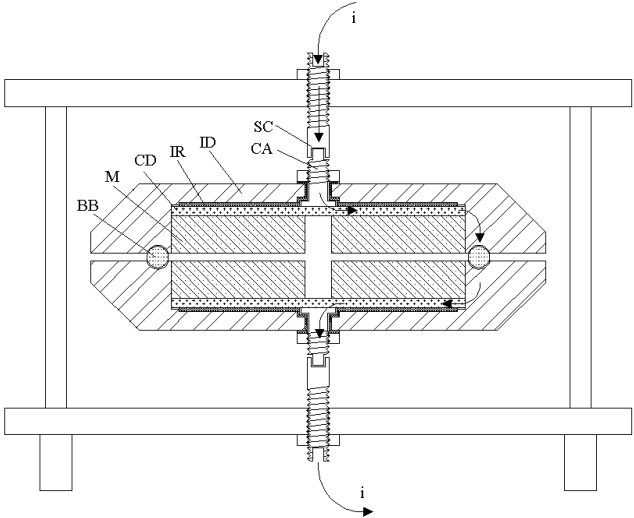

*Fig. 5. The twin Faraday disc. Two conducting discs replace the conducting hemispheres of the previous figure. The cylindrical magnet is cut by the equator in two equal parts and fixed to the discs. ID Iron Disc, IR Isolating Ring, CD Conducting Disc, M Magnet, BB Brass Balls, CA Conducting Axis, SC Sliding Contact.*

If we substitute a spiral for the magnet and conducting disc, we obtain a Twin Faraday Disc which works with continuous and alternate current.

## 5. The Conducting Spiral as an Electrodynamic Turbine or Universal Unipolar Generator.

In this section, for the sake of homogeneity and symmetry, we shall ignore the magnets and only consider the charges on the conductors and their relative movements.

Let us suppose that we substitute a circular current on the equator for the magnet enclosed in the sphere, Fig. 4. It is irrelevant whether the circular loop conductor rotates with one of the hemispheres or not. The circular current round the equator produces two equal and opposite torques on the radial or meridian currents (as demanded by the Angular Momentum Conservation Principle). The radial currents can not produce any torque in relation to the axis of the circular current (the forces are normal to the circular loop, and the torque in relation to the axis is null. Ampere's third experiment).[22]

The fact that the torque in relation to the axis of a circular loop current will always be null, makes it irrelevant whether the circular loop current is fixed or not to the radial current, as occurs with the magnet and Faraday's Disc. For this reason there is no conceptual difference between the disc and Faraday's revolving magnet, as is often stated.

"Faraday's disc should not be confused with the case of unipolar induction. In the latter the rotating disc is a magnet itself. This case is much complicated conceptually and never touched on in introductory physics courses."[23]

Therefore, it is [possible to connect circular and radial currents in series, thus forming a single current. In this way, we obtain a G-shape circuit which represents the Faraday Disc. With this simple line of thought in mind, I discovered the spiral to be a unipolar generator some forty years ago.[24] The G-shaped circuit is a most particular type of spiral. It may be asked whether any conducting spiral be a unipolar or acyclic generator similar to the Faraday Disc?

In fact it will, but it also constitutes a universal generator, which means it works with both continuous and alternate current. Further on, we shall see the difference between a logarithmic spiral and a G spiral.

A G–shaped loop or circuit is formed by a circular loop and a radial segment connected in series. A conducting spiral circuit is formed by a continuum of circular and radial elements. Fig. 6.

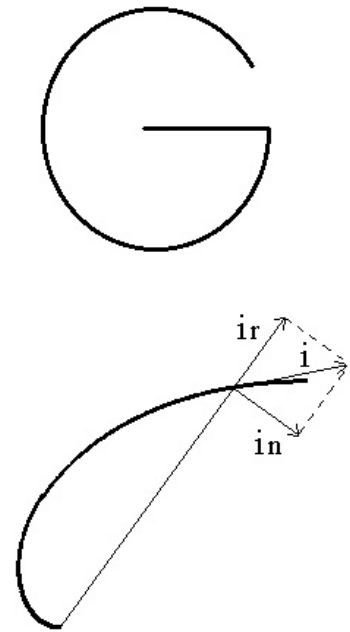

*Fig. 6 ( a) The "G" loop circuit formed by a circular loop and a radius. (b) In the spiral circuit, each element of the spiral can be decomposed into an element in the radius direction and another normal to the radius.*

For mechanical and electromagnetic experimental purposes, it is irrelevant whether the radial and circular currents belong or not to the same circuit. Either way, the resistance and inertia of the radial conductor (disc) are increased.

It is easy to prove experimentally that a spiral–shaped conductor is a unipolar or acyclic machine, whose rotation direction doesn't depend on the current direction. Such a machine was patented.[25] When I published my article on the conducting spiral,[26] I had not realized that it constituted an electromagnetic paradox which would afford a new vision of electromagnetic induction.

Every conducting spiral current element has two components, one radial, belonging to the radial current of the Faraday Disc, and another normal to the radius, belonging to the circular current or magnetic field of the magnet. When the current direction in the conducting spiral is changed, this implies changing simultaneously the current and the field direction. The torque and the emf generated in the spiral are proportional to $i^2$. The vector product of the radial and normal components of the charges' velocity defines a vector or vortex, which add together in the case of the spiral, producing a single perpendicular vector to the spiral plane.

While a continuous spiral, whether Archimedean or logarithmic, immediately suggests, by analogy with mechanics, an electrodynamic turbine; and that the electromagnetic torque is due to continuous rate of change of the electromagnetic angular moment of the current. Such is not the case with the G-spiral or the Faraday Disc.

According to this phenomenological description, the continuous variation of the rotation radius in a vortex of charges, produces the continuous variation in the angular electromagnetic moment of the current charges.

Magnetic moment $\mu = \pi r^2 i = \frac{1}{2} q v r$ is not really a momentum, properly speaking. This is because its units are not a momentum (Newton • s = Kg • m/s). The time rate of change of the magnetic moment is not a torque either.

$$\Gamma \neq d\mu /dt \qquad (2)$$

The definition of a magnetic moment is cinematic and does not take into account the forces that have generated it.

We shall see further on, that in this case what is really happening is the circulation of the electromagnetic momentum.

In the G-shaped circuit (the nearest equivalent to the Faraday Disc), the charges of the continuous current in the circular part (magnet) of the G loop, that produces the constant magnetic moment, on interacting with the charges of the radial current segment (disc), changes the electromagnetic angular moment of the current.

While some charges produce an electromagnetic angular moment when rotating on an axis, others produce the rate of change of the electromagnetic angular moment, as they interact with those that move away from or towards the axis.

The "G" loop produces a constant rate of change of electromagnetic angular moment (torque), in the same way as a continuous spiral. The torque is proportional to the magnetic moment of its circular part and the current of its straight part.

$$\Gamma \approx \mu I \qquad (3)$$

The interaction of the charges produces a vortex or "magnetron effect", or transformation of the electromagnetic lineal moment into angular moment.

Possibly, because this effect has never been observed in mechanics, the gravitation forces involved being too week, a comparison with the origin of electromagnetic induction has not been established. We shall come back to this point later.

Regarding the Lorentz Force, it has been stated that:

**"The emf is independent of the path in the conductor since only the radial components of the path elements contribute to the integral $\int(v \times B) \cdot dl$".[27]**

**This does not apply to the spiral, for as we may observe, the normal component to the radius generates a magnetic field which adds to the "external" magnetic field. In the case of the spiral it is superfluous. This fact constitutes an aspect of the spiral paradox.**

It flows from this, that the magnetic induction B, the torque and the emf, all depend on the path of the charges in the conductor.

Consequently, the unipolar emf and the current in the Faraday Disc are not independent of the way the disc current is closed.

**Does not the differential expression of the Lorentz Force applied to the calculation of the emf in the Faraday Disc not make more sense than the integral expression of Faraday's Law?**

We shall see the mining of both expressions from the point of view of the rate of change of the electromagnetic angular moment.

Later, the geometry of the conducting spiral will allow us to distinguish absolute unipolar emf and torque (due to the intrinsic or absolute rotation of the Faraday Disc or spiral) from the relative emf and torque, which depend at the same time on the spiral and the rest of the circuit.

In most cases where the Lorentz Force is used, the magnetic induction generated by the circuit to which the current i belongs is negligible compared to the external magnetic induction B

## 6. The "absolute – relative" duality in the spiral paradox.

Since the rotation direction of the conducting spiral doesn't depend on the current direction, this could induce us to believe that the spiral torque only depends on the direction of the winding, and that it does not depend on the external magnetic field that can be generated by the rest of the circuit that closes the spiral or by any other circuit. We can verify experimentally that this is not the case: the circuit that closes the spiral can increase, decrease, annul or change the spiral torque direction.

If the spiral is closed by a radial conductor, its torque is "absolute", that is to say it is specific to the spiral and does not depend on the rest of the circuit, and for this reason does not increase or diminish. The spiral's torque does not depend on the radial conductor, however, the torque in the radial conductor depends on the spiral. By contrast, if the spiral is closed by a curved conductor or by another spiral, their torques are dependent or "relative".

The geometry of the spiral allows us to see this paradox or duality of unipolar induction clearly.

While the radial conductor neither increases nor diminishes the torque of the spiral, we cannot generate this torque without this conductor, which closes the circuit and satisfies the principle of the conservation of the angular moment.

This absolute-relative duality of torque and unipolar induction suggests that the elctromagnetic mass, as also the machanical mass is inert and gravitational. The word electromagnetic expreses the gravitational-inert duality of the charge.

The conducting spiral constitutes the most simple and beautifull form of the continuous transformation of electric into mechanical energy. This happens through the continuous variation of the lineal and angular moments of the electromagnetic field and matter.

## 7. The conservation of angular moments in a circuit.

When we state that a continuous current in a circular loop generates a constant magnetic moment, similar to the orbital magnetic moment of an atom, there is an electric field normal to the charges' trajectory in both cases. In the case of the circular loop, this electric field is generated by the normal constraint forces of the conductor. For this reason, when we mention a magnetic moment, the electric field is always normal to the charges' velocity. When a component of an electric field in the direction of the charges' velocity exists, a rate of change of the electromagnetic moment is produced. Fig.7.

If we superimpose a normal electric field on a magnetic field, a Poynting's vector circulation is generated.

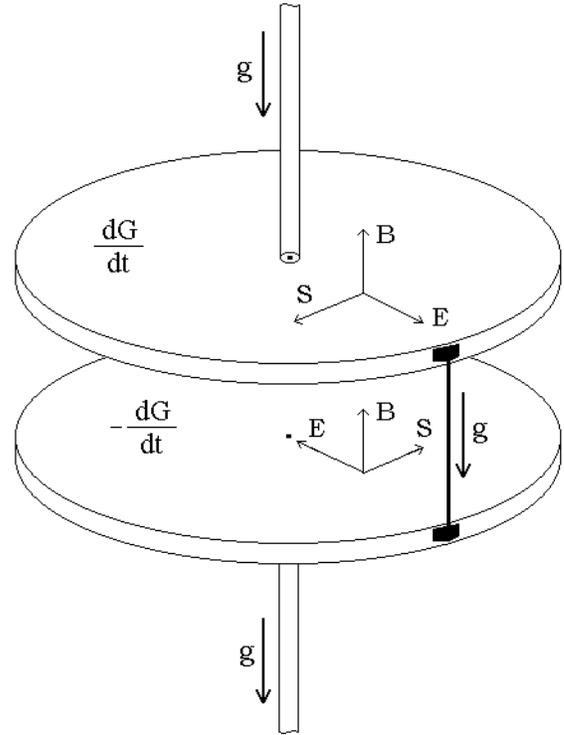

*Fig.7. The twin Faraday Disc. Magnetic induction B perpendicular to the discs transforms the continuous radial circulation of the electromagnetic lineal momentum p, in in two constant time rates of change of the electromagnetic angular momentum dL/dt.*

Electric charges moving in the direction of the radial electric field on the discs, produce two time rates of change of the angular moment of the electromagnetic field and angular moment of matter, giving rise to equal and opposite torques as demanded by the principle of the conservation of angular momentum. Reciprocally, the rate of change of the angular moments of matter produces the rate of change of angular electromagnetic moment or emf.

The Faraday Disc and the conducting spiral constitute experimental proof of this.

We shall now see how magnetic moment or electromagnetic angular momentum and its rate of change is generated in the conducting spiral.

A conducting spiral, just as the Faraday Disc, is an open circuit. If we were to form a closed circuit placing face to face two identical spirals (rotor - stator), the normal components of their currents would be oriented in the same direction and their magnetic fields would add on to each other, however, the radial components of the current elements are oriented in opposite directions. So, while the sum of the normal components of the current that generate the magnetic moment can have any value, the sum of the radial components of the current will always be zero in all closed circuits. See Fig. 8.

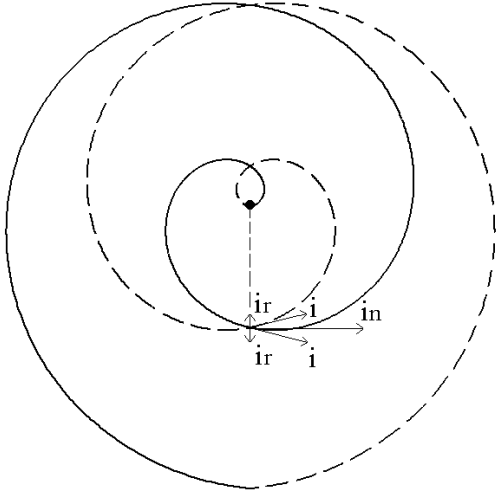

*Fig. 8. A closed circuit formed by two spirals of equal step and opposite direction. The normal components to the radius do not cancel each other out.*

From this, we may deduce, as is well known, that the rate of change of the moment of the charges of a continuous current will be zero in any closed circuit.

The fact that the sum of the radial components is always zero, divides the circuit in two, and establishes the equality of the torques *and of the emfs*. To each radial element $i_r$, there corresponds another, at the same distance from the axis of rotation and which produces an equal and opposite torque. This is because the electromagnetic forces are normal to the conductors. The normal elements $i_n$ that generate the constant magnetic moment of the current charges can increase or diminish the rate of change of angular moment (torques) according to their direction, generated by the radial components. They are equal because of their geometric origin.

When in a rigid circuit made up of one or two spirals there circulates a continuous current, two continuous equal and opposite rates of change of the magnetic and mechanical moments are generated, which produce two equal and opposite torques.

Reciprocally, when the Faraday Disc or the conducting spiral is made to rotate with a constant angular velocity, (a non-rigid circuit) the normal constraint forces of the conductor propel the charges generating a constant current which produces two constant, equal and opposite rates of change of the electromagnetic angular moment or emfs.

When a torque acts on the Faraday Disc or conducting spiral, at the same time as the angular velocity ω increases, the intensity of the current and magnetic flux density also increase, which implies an increase in the time rate of change of the electromagnetic angular moment. Reciprocally, if we vary the current intensity, the magnetic flux density will also vary in the same way as the time rates of change of the angular moments of the electromagnetic field and matter.

The orbital angular moment of a charge depends on the velocity magnitude and on the radius, $L \approx v^2/r$

When the conduction electrons move throught a conducting spiral with constant drift speed, the current intensity and the magnetic flux density are constant. However, a constant rate of change of the angular moment of the electrons is produced, due to the variation of the trajectory radius, which gives rise to constant electromagnetic torque of the spiral and emf.

$$\varepsilon = -dL/dt = -\Gamma \qquad (4)$$

The variation in emf and current intensity implies a variation in magnetic flux density, the drift speed of the electrons, the electromagnetic angular moment and the spiral's torque.

$$d\varepsilon/dt = -d\phi/dt = -d\Gamma/dt \qquad (5)$$

where $\phi$ is the magnetic flux density, $\varepsilon$, $L$ and $\Gamma$ are the emf, the electromagnetic angular moment and the torque respectively.

The generation of constant emf and current are due to two constant time rates of change of electromagnetic angular moment of the current, which are equal and opposite respectively in each of the two parts of the circuit that move in relation to each other.

According to Faraday's induction Law $\varepsilon = -d\Phi/dt$, ($\Phi$ is the magnetic flux) the generation of a constant emf and current is due to a constant time rate of change in the magnetic flux; as we might incorrectly conclude from the following experiment, the generation of a constant emf through the deformation of a circuit in a uniform magnetic field (a rectangular loop in which one of the sides moves at a constant velocity) the magnetic flux enclosed in a circuit may vary, but its density will remain constant.

The generation of a constant emf necessarily implies a constant density of the magnetic flux.

On producing a deformation in a circuit, two equal and opposite variations of the electromagnetic angular moment of the current can be produced; even though the density of the magnetic flux remains unchanged.

In the symmetrical Faray Disc we saw that the two emfs must be *opposite* (in the same way as angular momenta) for otherwise they will cancel each other out. These two emfs or constant time rates of change of the angular moment of the electromagnetic field, are generated respectively in each of the parts in relative movement. Here the problem of finding the seat of the emf, as mentioned in the introduction, is also solved.

The emfs are just as locatable as the torques.

While the origen of the two emfs is in the rotating spiral (as the Lorentz Force has it), they are both the result of the interaction between the two spirals or parts of the circuit. Let us consider an example, by way of analogy, from the point of view of mechanics. If two bodies are at rest and a force acts on one of them causing it to collide with the other, we can state that the body responsible for the collision is the body that was impelled. However as a result of the collision, two equal and opposite moments are generated in the interacting bodies.

Now, taking into account these ideas we are in a position to understand the Faraday Disc and electromagnetic induction.

To summarize, electromagnetic induction is due to the coexistence and conservation of the angular moment of the electromagnetic field and angular moment of matter.

For this reason a closed loop in the primary part of a transformer diminishes induction in the secondary. Lenz Law and the diamagnetism of superconductors constitute the most perfect and constant demonstration that electromagnetic induction is due to the conservation of the electromagnetic angular moment of the current charges.

We shall demonstrate, taking another aspect of the spiral paradox, that electromagnetic torque is linked to the time rate of change of the angular moments of the electromagnetic field and matter.

## 8. The paradox of unipolar torque in the spiral.

If the circuit that closes the spiral is very short and the only magnetic field is that of the spiral itself, we should expect that while the spiral's step diminishes and the number of turns increases (assuming the current's intensity and maximum radius are constant), its torque as well as its magnetic moment will increase.

Surprisingly, this is not the case. While the resistance and heat generated in the spiral increase rapidly with the number of turns, its torque does not change perceptibly.

Spirals of 19 cm. diameter were made having different steps. The copper wire used was 3 mm. diameter.

In order to facilitate rotation, the center and exterior electrical contacts to the spirals were made through the medium of mercury. Spirals and mercury were enclosed in hermetically–tight, transparent Lucite boxes. Fig. 9.

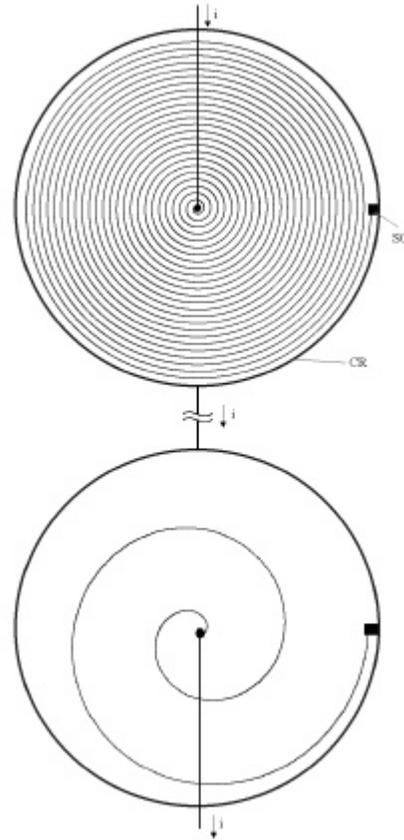

*Fig. 9 The spiral paradox. Conducting spirals of different step values and equal radius maximum connected in series and sufficiently far apart so that the interaction of their fields would be negligible. The spiral's torque increases very slowly with the number of turns, which is the contrary of what we would expect. CR Collecting Ring, SC Sliding Contact.*

In this experiment, two spirals of the same diameter (19 cm) and widely different step (3 and 100 mm), their centers 1,5 m apart, were connected in series.

**If current, and maximum radius are constant we should expect the torque of the spiral to increase with the number of turns, as the magnetic moment**

**increases. Quite to the contrary, torque is not found to vary perceptibly with the number or turns.**

To the extent that the step value of the spiral diminishes, the number of turns and the magnetic moment of the current charges increase, but this implies, necessarily, that the rate of change of the angular magnetic moment of the current diminishes.

The most logical and simplest explanation for this paradox consists in attributing the electromagnetic torque to the time rate of change of the electromagnetic angular moment of the current.

$$\Gamma = dL/dt \qquad (6)$$

We have seen how the electromagnetic torque of the spiral conductor depends on the product of its two components (radial and normal) of the charges' velocity.

Given that the speed of the charges' drift is constant all along the spiral conductor, the angular moment changes. By contrast to what happens in the elliptical trajectory of planets, the angular velocity changes in order to maintain the angular moment constant (Kepler's Second Law).

At the same time as the spiral step diminishes and the number of turns increases, the normal component of the charges' velocity and the magnetic moment $\mu$ (or the angular moment of the magnetic field) increases, but the radial component of the velocity diminishes, as also the time rate of change of the angular moments of the electromagnetic field and matter.

That is to say that with the spiral, the normal velocity of the charges increases, as the radial velocity disminishes.

The increment in magnetic induction and in magnetic moment of the current's charges, when increasing the number of turns in the spiral lowers the rate of change of the turn radius of the charges and the rate of change of the moment. When the spiral step tends to zero or infinity the torque tends to zero, too.

Furthermore, the sensitivity of the spiral to self-torque depends on the relative rate of change of the angular moment $\Delta L/L$. For the increase in the spiral's moment implies an increase in weight, friction–torque and resistance.

A multi-layered cylindrical coil is equivalent to a spiral in which the relative rate of change of moment is very small, this is due to its magnetic moment being very large, and its rate of change very small. This is the reason why the rotation of a cylindrical coil round its axis is extremely difficult to bring about. No cases have been reported.

If the radial $i_r$ and normal $i_n$ components of the current, in a conducting spiral contribute to the same extent to its electromagnetic torque and this being due to the product of the components, torque will be at a maximum when components are equal.

Archimedes' spiral has a constant step $r = k\theta$ and does not comply with the above condition because the radial component diminishes as the normal component increases with the widening of the spiral. The spiral on which all points have equal normal and radial current components, $i_r = i_n$ will be that which has maximum torque.

If this condition is fulfilled, a logarithmic spiral is obtained:

$$rd\theta = dr \quad d\theta = dr/r \quad \theta = \ln r \quad r = e^\theta \qquad (7)$$

Where the dimensional constant $k = 1$

**The logarithmic spiral is the shortest spiral with the maximum electromagnetic torque.**

This can easily be verified by measuring the torques with a torsion balance of, for example, a G-shaped circuit and logarithmic spiral of equal length. Fig. 10.

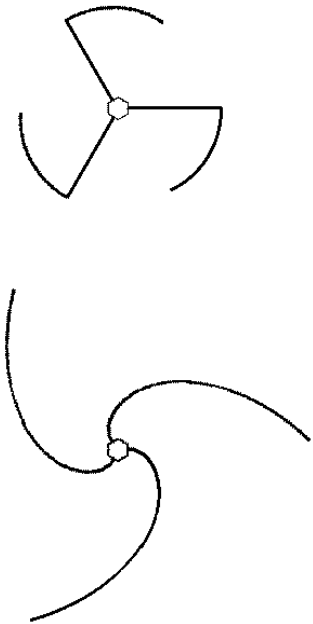

*Fig. 10. Two three–arm spirals of equal length but different shape (G and logarithmic). The mobile contacts at the center and ends of the spirals were made through mercury and closed by a radial current so as not to change their self- torque.*

The deformation of a circuit by electromagnetic forces tends to diminish the time rate of change of the electromagnetic angular moment. This means the deformation of the circuit will tend to conserve the electromagnetic angular moment.

## 9. Mechanical mass and Electromagnetic mass.

In clasical mechanics a distinction is made between inert and gravitational mass as two aspeccts of matter whose nature is totally diferent and which are defined and measured by their differents effects. Apparently, the charge also has a dual, inert and gravitational nature. This is shown by the fact that the charge units in the mks system (coul), and the Gauss system (statcoul) are physically different, and are measured by their different effects. The constant $1/4\pi\varepsilon_0$ $3x10^9$ by which they are related, has dimensions.

It is usual not to express the dimensions of the constant $3x10^9$ and to take the relation between coul and statcoul as simply a change of scale, analogous to m and cm in these unit measurement systems. While in the mks system, the charge (coul) would be measured according to its inert nature, in the Gauss system (statcoul) this would be done according to its gravitational nature.

It is said that the electron has a mass, and the relationship between its charge and mass is $e/m_e = 1.76x10^{11}$ coul/Kg. If we attribute inertia of charge to the fact it has a mechanical mass: what will the attributes of the elctromagnetic mass of the charge be?

"Suppose an electron is moving at a uniform velocity through space, assuming for a moment that the velocity is low compared with the speed of light. Associated with this moving electron, there is a momentum – even if the electron had no mass before it was charged - because of the momentum in the electromagnetic field".[28]

The electromagnetic mass of an electron $m_{elec} = 2e^2/3ac^3$ is calculated from the density of the lineal moment of the electromagnetic field $g = \varepsilon_0$ E x B integrated in all the space (corresponding to the inert part) and to the energy of the electrostatic field $U = q^2 / 8\pi\varepsilon_0$ (corresponding to the gravitational part), $r_0 = e^2 / m_{elec}c_2 = 3/2a$ which is called the classic radius of an electron.[29]

Electromagnetic mass thus calculated is just as inert and gravitational as mechanical mass. They coexist in the same way as the duality particle-field or, mass and energy. They are the two sides of a coin.

Are the inert and gravitational nature of the charge independent? Can we measure them separately, as as we apparently do with mechanical mass?

When we say that inert mechanical mass is intrinsic, and that it does not depend on the presence of others masses, what does this mean? Can we really see and measure the inertia of a mass without the presence of another mass? It is possible to apply a force to a mass withhout the presence of another mass?.

The existence of charges of opposite sign, whose inertia is very small and gravity very strong allows us to change the proportion very quickly, generating very intense fields.

Contrariwise, the gravitational nature of mechanical mass is weak as compared to its inertia, which makes it impossible to produce a significant and quick change in the gravitational field (as for example, measuring the speed of propagation of the gravitational field)

In the following section we shall see that electromagnetic torque by a apparently continuous current in a conducting spiral makes the double, gravitational and inert nature of the electromagnetic mass clearly apparent.

## 10. The spiral's torque and electromagnetic mass.

In section 7, I suggested the hypothesis that the radial and normal components of a current contribute equally to the electromagnetic torque of a spiral. In justifying the hypothesis, it will be seen that this is equivalent to attributing the electromagnetic torque to the time rate of change of the electromagnetic angular moment of the current's charges in the spiral. Also, the auto- torque of the spiral does not depend on ineraction with the rest of the circuit when this consists of a radial conductor. For this to happen, the electromagnetic auto-torque generated in a logarithmic spiral must be the measure for a kind of conducting electron mass circulating in the spiral.

Given a logarithmic spiral $r = e^{k\theta}$ with a constant $k = 1$. If the speed of displacement of the electrons $v_d$ is constant, the modules of the radial and normal components will be equal and constant throughout a logarithmic spiral of constant 1. Fig. 11.

According to the Biot-Savart Law,

$$dB=\mu_0/4\pi \; dl \times r/r^3 \qquad (8)$$

dlxr is the normal component of the current element di, that generates the angular moment or the magnetic field at the point r.

The contribution to the variation of the spiral's angular moment $d\Gamma = dl/dt$ or torque, produced by an element of the current di, depends on the magnetic field. This magnetic field is produced by all the normal components to it, of the spiral's and the rest of the circuit's current elements.

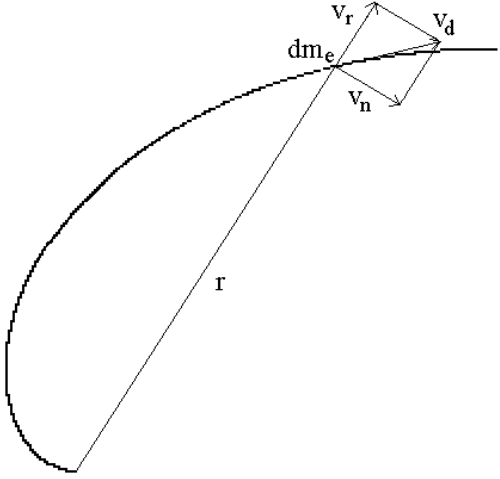

*Fig. 11. The electromagnetic auto-torque generating by the conducting electrons in a logarithmic spiral is a measure of the inertia of electrons, which depends on the shape of the circuit.*

If the magnetic field produced by a circuit which closes a spiral is null in relation to the spiral center and the magnetic induction B along its length varies inversely proportionally to its radius, all the currents elements of a logarithmic spiral should contribute equally to its torque and make it possible to define the constant inert mass of the conducting electrons $m_e$ in a way analogous to mechanical mass.

That is to say, the angular moment dl produced by a $dm_e$ would be $dl = r\, v_n\, dm_e$ and its contribution to the electromagnetic torque

$$d\Gamma = \frac{dl}{dt} = \frac{dr}{dt} v_n dm_e = v_r v_n dm_e \qquad (9)$$

The variation in angular moment dl of the $dm_e$ of the conducting electrons would be constant all along the logarithmic spiral. We may also ascertain that the variation is maximum in a logarithmic spiral of constant 1, as shown in the previous section, that is to say,

$$|v_r| = |v_n| = \frac{\sqrt{2}}{2}|v_d|$$

$$\Gamma = \int d\Gamma = \frac{v_d^2}{2}\int dm_e = \frac{v_d^2}{2} M_e \qquad (10)$$

in which Γ is the auto-torque of the spiral, and $M_e$ is the electromagnetic mass of all the conducting electrons contained in the logarithmic spiral.

When the circuit closing the spiral is not radial, but long and curved, the interaction of the spiral's current elements with the normal components of the current elements in the rest of the circuit which contributes to the total angular moment (magnetic field) causing the spiral's torque to change.

When we change the circuit shape which closes the spiral, keeping the current constant, this is equivalent, by analogy with mechanics, to varying the inertia of the conducting electrons in the spiral.

When we measure the magnetic field generated by a current that is equivalent to measuring the inertia of the conducting electrons for a specific value of its interaction energy and velocity.

The electromagnetic torque of the spiral will allow us to see that gravitational and inert nature of the electromagnetic mass are not independent, as happens with mechanical mass in the General Theory of Relativity in which mass depends on its velocity and its position energy.

As a result of this, the conducting spiral constitutes a simple and beautiful example of the intimate relation between mechanics and electromagnetism.

## 11. Electromagnetic induction and the time rate of change of angular moment.

The angular electromagnetic moment generated by the circulation of the charges (just as the angular moment of a particle) depends on its angular velocity and its rotational radius

$$L = I \cdot \omega \qquad (11)$$

Because of this, there are three possible ways of varying the electromagnetic angular moment of a charge. These three ways give rise to the three known forms of electromagnetic induction:

a) Variation of the angular velocity magnitude: alternating currents, transformers.

b) Variation of the angular velocity direction: Alternating current machines or cyclic machines.

c) Variation of the rotation radius: unipolar induction, direct current machines or acyclic machines.

Unipolar induction ceases to be an exception and now confirm the new induction law.

## 12. The normal constraint forces to the conductor and magnetic field.

For the moving charges in the conducting spiral, the only "magnetic field" present is that generated by the charges constraint forces normal to the conductor, which change the charges' velocity and produce the electromagnetic angular moment.

This suggests there exists an inversion of cause and effect when defining the magnetic force on a current.

**THE MAGNETIC FORCE ON A CURRENT. "Because a magnetic field exerts a sideways force on a moving charge, it should also exert a sideways force on wire carrying a current."[30]**

The conducting spiral proves that the normal constraint forces on the charges, which circulating in the conductor, produce the electromagnetic angular moment and its variation. The variation of the electromagnetic angular moment induces an opposite moment in such a away as to conserve the electromagnetic angular moment. **In this way, the reaction of the normal constraint forces gets transmitted.** For this reason, the direction of rotation of the spiral depends on the total moment resulting from all the charges, whether within or outside the spiral.

The constraint forces of the charges, as also the centripetal force of a particle, demonstrate the inert nature of a charge.

The constraint forces of the charges, which are normal to the conductors, generate the magnetic field, which in turn generates the normal forces to de conductors. By analogy to what happens in mechanics: the centripetal forces gives rise to the inertia of the particle, which in turn generates the centrifugal force.

The magnetic field (relativistic effect) and magnetic force, is due to the inert property of the charge.

The Lorentz equation $F = q(E+v \times B)$ expresses the double inert and gravitation nature of the charge.

Without the constrain forces to the charges on a conductor it is also impossible to generate and/or measure an electric field. However, these forces are not explicit in Maxwell's equations.

The fact that we are able to define and calculate, for example, the angular moment of a particle without taking into account a centripetal constrain force, does not imply that we can generate or explain the angular moment of a particle without this force.

As a result of this, we can't explain the generation of the angular electromagnetic moment and of the angular moment of matter by the circulation of the charges without taking into account the charges constraint forces normal to the conductors. Without them it is not possible to generate a magnetic field or the "mystic circulating flow of energy"[31] produced by the circulation of charges. This circulation necessarily implies an electric field normal to the trajectory. This field is produced by some kind of constraint force. Apparently, Rowland's experiment confirm this. However, the question can be asked whether a reciprocal experiment would be meaningful. Could the rotation of the compass in place of the charged disc detect a magnetic field?.

## 13. Angular moment variation and mass induction.

The analogies between mass and charge in experiments carried out on conducting spirals suggest the existence of a very weak induction of mass or mechanical Lorentz Force. Moreover, General Relativity Theory, which shows that mass depends on velocity and on position energy, should be borne in mind.

In a continuous spiral, the velocity of particles circulating in it have two components, radial and normal respect to the spiral's center, and their product gives rise to the variation in angular moment of the particles. In the sections 7 and 9 we saw that the logarithmic spiral of constant 1, constitutes the turbine with maximum torque per unit length, both for charge and mass.

For a particle of mass moving at constant speed to experience a continuous variation in angular moment in relation to a point, its velocity must have two components, one radial and other normal in relation to this point as happens in a spiral. In the specific case of the logarithmic spiral, when traveled over by a particle with a constant speed, the magnitude of the normal and radial components of the speed are constant throughout the spiral producing a constant variation of the angular moment on the particle in relation to spiral's center. On the other hand, a set of two charges may undergo a variation of its angular moment in relation to a point if the speed of one of them is radial and the other normal with regard to that point. **This suggests that the difference is due to the strong electric force among the charges, compared to a weak gravitational interaction between the masses.**

Torque or variation of unipolar electromagnetic angular moment depends on the product of the radial and normal currents, which do not need to form part necessarily, of the same circuit. We could generate a small radial current and a large normal or circular current or vice versa, with two independent circuits. There is no parallel for this in classical mechanics.

**The logarithmic spiral and the G spiral (Faraday Disc) as turbines symbolize the difference between mass and charge.**

While in the case of the G spiral, as a hydraulic turbine, the connexion between the radial part and circular part is fundamental in order to produce the variation in angular moment in the mass at the connexion point, this is irrelevant in the electrodynamic turbine, where the variation in angular moment of the charge occurs all along the radius.

According to this analogy, the Foucault Pendulum is equivalent to a mechanical Faraday Disc and should describe a small oscillation perpendicular to its plane, which would form an extremely eccentric ellipse.

In a future article, some experiments to detect this very weak induction in mass, using mechanical resonance, will be suggested.

In spite of the small probability that these experiments be carried out successfully, I believe it is worth trying. I remember how some colleagues made fun of me when I set out to make the spiral revolve forty years ago.

A preliminary summary of this study was presented at the First Venezuelan Congress of Physics, Mérida, Venezuela, December 1997.

I should like to thank my friend and colleague Gianfranco Spavieri for the useful conversations on the subject of this paper.

I should also like to thank Antonio Albornoz of the faculty laboratory, for his invaluable help in building the experiment apparatus.

My thanks also go to my friend Vincent Morley who helped me check the English translation of the text.

This work is financed by CDCHT (Commission for Scientific, Humanistic and Technological Development), Universidad de Los Andes, Venezuela.

For further information on de subject and the reasons for my article being rejected by Am. J. Phys. I include some commentaries by the referee and editor together with my letter which remains unanswered.

November 10, 1999

Prof. Albert Serra-Valls
Depto. de Fisica, Fac. de Ciencias
Universidad de Los Andes
Zona Postal 5101
Merida
VENEZUELA

Dear Prof. Serra-Valls:

I write in connection with your manuscript (MS #10396), "Electromagnetic Induction and the Conservation of Momentum of the Charges in the Spiral Paradox".

I enclose a referee report on the manuscript, and having studied this report and having reread the manuscript myself, I agree with the conclusion of the referee that this paper should not be published in the American Journal of Physics.

It is now time to end editorial consideration of this manuscript for this Journal; you are free now to submit the manuscript elsewhere, but we will not consider further revisions of this manuscript.

Sincerely yours,

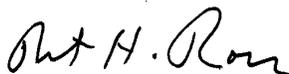

Robert H. Romer

RHR/kk
Enclosure
cc: Referee

Author: A. Serra-Valls

Title: "Electromagnetic Induction and the Conservation of Momentum of the Charges in the Spiral Paradox"

1. Briefly summarize why you believe this manuscript would or would not be of interest and/or value to the readers of AJP.

in AJP
The recent publications of an exchange between Galili, Kaplan and Scorgie indicates that there is still a controversy about unipolar induction + Faraday's law. I believe that Scorgie rightfully has the "last word" on this subject. A manuscript which successfully criticizes S's publications (in AJP + Eur. J. Phys) would be exciting.

2. Is the manuscript technically correct?

No, pls see report

3. Are the style, grammar, level, etc. of this manuscript suitable for publication?

Grammar + level are fine; style is too diffuse — pls see report

4. Are the references to previous work, in this journal and elsewhere, adequate? If not, please elaborate.

Yes.

5. Overall recommendation:
____ Enthusiastically recommend publication
____ Recommend publication
__X__ Recommend against publication

The majority of manuscripts can profit from revision. If you have recommended publication of this paper, please advise us about the need for revision:

____ Acceptable in present form
____ Acceptable if revised

____ I should see the revision
____ I do not need to see the revision

Referee's Report on

"Electromagnetic Induction and the Conservation of Momentum of
the Charges in the Spiral Paradox"

by Albert Serra-Valls

As one of its aims, the American Journal of Physics seeks to promote discussion
of controversies, while minimizing diffuse argumentation. I have tried
hard to see this paper as doing the former, but have not succeeded.

Some controversy does surround electromagnetic induction. In recent AJP
publications, Galili & Kaplan argue that that the integral form of Faraday's
law is neither of "explanatory nor of general power".
Scorgie challenges G & K in a short note.

The present manuscript introduces the controversy and extends it to
include Richard Feynman and Dale Corson. This is done in a single key
paragraph in the middle of page 2.
Unfortunately, the introduction diffuses the conflicts so much that
NO short article can resolve the issues.

For instance, the author states that "[Scorgie] questions the validity of
the Lorentz force". If Scorgie really believed that, he could hardly have
convinced others of it in a single page. After all, most physicists believe
that the E & B fields are defined by the Lorentz force law. Actually,
S merely states that Faraday's law is needed because, even in an object as
common as the dynamo, we don't have enough information to use Lorentz's law.

The author also misses the mark when he attacks Feynman at the bottom of
page 3. His argument there seems to be basically with Figure 4 (which is
copied directly from Feynman's classic textbook). No one doubts that the
illustrated generator will work; it is just not very efficient.

At the top of the next page, the author appears to attack Corson for
doing what all textbook writers do: namely take the changing flux in
the Faraday disk generator to be that swept by the radius of the disk.
Alternatively, the author does not believe Galili and Kaplan when they
write (in AJP) that "Faraday's law does not account for Faraday's
generator".

I don't believe G & K either. But that is not the point. The difficulty
with the present manuscript is that by diffusing the present situation
so badly, the author kills the interest of the reader in understanding
the author's new contributions - his experiment and the spiral paradox.
This reader's annoyance is not helped by the author's unconventional use
of the expression "the magnetic moment of the charges" in the abstract
and text. The magnetic moment of the charges is the magnetic moment of the
electron. I think the author means the magnetic moment of the currents,
but I am not sure.

I do not believe the present manuscript is publishable without major
revision. I suggest that the author place his work in the context
of current textbooks such as Feynman's or Lorraine and Corson's.
He also should strive to make one paragraph follow logically from
its predecessor. Extraneous material such as the single paragraph
on Arago's disk or the second paragraph on page 4 make the
manuscript very hard for any reader to follow.

The Editor   Albert Serra - Valls
American Journal of Physics   Apartado Postal 630,
Merrill Science Building, Room 222   Mérida, 5101
Box 2262   Venezuela.
Amherst College   e-mail: serra@cantv.net
Amherst, Massachusetts 01002
Mérida, December, 13 - 1999

Dear Mr. Romer,

I hope you received my letter of Nov. 9$^{th}$ with the manuscript of the new version of my paper, which must have been in the post at the same time as your letter of Nov. 10$^{th}$ in which you inform me your rejection of the manuscript I sent on June 30$^{th}$.

It is a pity that my latest manuscript did not reach you in time, for I feel sure that the referee would have viewed it more favorably.

Unfortunately, There are some mistakes in the first manuscript, as at the time I still thought that the two equal and opposite variations of the electromagnetic moment generated by the charges of the current in the closed circuit of the Faraday Disc saved Faraday's Induction Law.

However, quite on the contrary, as you may see, my experiments with the symmetrical Faraday Disc and the conducting spiral confirm (agreeing with Feynman, Galili and Kaplan) that unipolar induction does not only constitute an exception to Faraday's Induction Law, but that this law is wrong.

On carefully rereading "Electromagnetic Induction in Deformable Circuits" I realize I should not have included Scorgie among those who question the Lorentz force, as the referee quite correctly pointed out.

However, the title "Only the integral form of the law of electromagnetic induction explains the dynamo" as well as the text of the brief article published in Am. J. Phys. leads one to believe that Mr. Scorgie has objections with regard to the Lorentz force.

I accept that I made some mistakes and consequently had my article rejected. What I cannot accept is the following remark "The author also misses the mark when he attacks Feynman at the bottom of the page 3. His argument there seems to be basically with Figure 4 (which is copied directly from Feynman's classic textbook)". Nothing could be further from the truth, for not only do I consider Mr. Feynman to be among the best contemporary physicians and pedagogues but that without his pointing the way, quite possibly I should never have written my article. I am quite sure that if the referee should read the whole of my article posted on Nov. 9$^{th}$ carefully, he would see this. Precisely the line of thought in this article is based on the coexistence of the angular moment of the electromagnetic field and the angular moment of matter and its conservation, I owe this to Mr. Feynman (Lectures on Physics, Chapter 27-5) as can be seen in Figure 8 and reference 26. The reason for reproducing the Figure of the disc in Feynman's book (probably not drawn by the author himself) is to show how it differs from the Barlow Wheel and from Faraday's Disc and rotary magnet, Figure 1, in which the field rotational symmetry axis of the magnetic field of the cylindrical magnet coincides with the rotation axis. In cases where the axes do not coincide, Eddy currents are induced in the discs. Scorgie also mentions this towards the end of his article in the Eur. J. Phys. With the same aim in mind, I mention Arago's Disc in the introduction. For the referee, this pedagogical and historical reference appears estrange and incomprehensible for he writes: "Extraneous material such as the single paragraph on Arago's disc or the second paragraph on page 4 make the manuscript very hard for any reader to follow."

When I mention "the magnetic moment of the charges", I have in mind, as the referee suggests, the magnetic moment of the currents. In my most recent manuscript of Nov. 9$^{th}$ I clarify the difference between the magnetic moment of a constant current and the electromagnetic angular moment of a current.

As may be appreciated in my revised manuscript, I share Feynman, Galili and Kaplan's opinions that "Faraday's Law does not account for Faraday's generator." So I hope that for holding this opinion I will not be accused of attacking Faraday who I admire very much. However, in science, as is well known, no one has the last word.

I have made a concentrate effort to cut to a minimum diffuse argument and to be as to the point as possible. Yet, intuition is diffuse and deceptive by nature, while being responsible for every discovery not produced by chance.

**I consider that the referee is not being consistent when he says, for instance, "Some controversy does surround electromagnetic induction. In recent AJP publications, Galili & Kaplan argue that that the integral form of Faraday's law is neither of "explanatory nor of general power". Scorgie challenges G&K in a short note. I don't believe G&K either. But that is not the point. The difficulty with the present manuscript is that by diffusing the present situation so badly, the author kills the interest of the reader in understanding the author's new contributions - his experiment and the spiral paradox."**

Apparently, the referee is justifying his not heaving read my manuscript (the first reading of a new approach is not easy going) and this is shown in his not giving any opinion on the most relevant and important points based on irrefutable experimental facts, namely:

a) Unipolar emf and torque are due to the constant rate of change of the angular moment of the electromagnetic field and matter, which coexist. This constitutes a new analogy between mechanics and electromagnetism.

b) The different forms of electromagnetic induction are due to the three possible ways of varying the electromagnetic angular moment of the current's charges; these three ways correspond to the different and "diffuse" ways of varying the magnetic flux mentioned by Scorgie in the Eur. J. Phys.

c) The spiral paradox proves that constant unipolar torque is due to the constant rate of change of the electromagnetic angular moment of the current's charges.

d) The conducting spiral proves that unipolar induction is produced by a charges vortex; which is to say that the Faraday Disc constitute an electrodynamic turbine.

e) The curious "absolute-relative" duality of unipolar induction, at the origin of controversies, apparently so "diffuse", becomes clear and manifest in the conducting spiral.

The fact that the referee does not mention any of these most specific and important points is no compatible with the statement that "The American Journal of Physics seeks to promote discussion of controversies".

The referee does not object to the content of the article or to the experiments. It would seem that his objections are more to the style and perhaps for this reason he does not reject my article outright, to quote from his comments: " I do not believe the present manuscript is publishable without major revision"

I would like to take you up on this and suggest having the opinion of a second referee.

In 1969-1970 an Am. J. Phys. referee twice rejected my manuscript giving as his reason that the conducting spiral was a "mind experiment" which could not really revolve. After a further examination of my article it was published in the Am. J. Phys. (A. Serra - Valls and C. Gago - Bousquet, "Conducting Spiral as an Acyclic or Unipolar Machine", Am. J. Phys. Vol.38, N.11, pp.1273-1276, Nov. 1970)

It is quite possible that my manuscript might contains some minor error which could easily be put right, but it should obvious that this hypothetical possibility should not obscure the results of a long study and much experimental work, no less than that which went into my earlier article which you published and to which my new article is the continuation, which is why I think my new article should appear in the Am. J. Phys. too.

You will excuse my insisting, but had I not done so on the occasion of my first article you would never have published it. Perseverance is certainly a virtue in science.

Sincerely yours,

Prof. Albert Serra - Valls
Dep. De Física, Fac. de Ciencias
Universidad de Los Andes
Manuscipt number: 10396